\documentclass[superscriptaddress,aps,pra,twocolumn,showpacs,nofootinbib, longbibliography, notitlepage,floatfix]{revtex4-2}
\usepackage{etex}
\usepackage{amsmath,amssymb,amsthm}
\usepackage[colorlinks=true,citecolor=blue,urlcolor=blue]{hyperref}
\usepackage[pdftex]{graphicx}
\usepackage{cleveref}
\usepackage{times,txfonts}
\usepackage{braket}
\usepackage{color}
\usepackage{natbib}
\usepackage{bm}
\usepackage{soul}
\usepackage{amsmath,blkarray}
\usepackage{mathtools}
\usepackage{latexsym}
\usepackage{tabularx, booktabs}
\usepackage{graphics,epstopdf}
\usepackage{graphicx}
\usepackage{float}
\usepackage{amsfonts}
\usepackage{tikz}
\usepackage{wrapfig}
\usepackage{upgreek}
\usetikzlibrary{quantikz}
\usepackage{color,soul}
\usepackage[T1]{fontenc}

\newcommand{\be}{\begin{equation}}
\newcommand{\ee}{\end{equation}}
\newcommand{\ba}{\begin{eqnarray}}
\newcommand{\ea}{\end{eqnarray}}

\usepackage{multirow}
\usepackage{appendix}
\usepackage{url}
\usepackage{listings}

\usepackage{comment}

\begin{document}
\title{Digitized Counterdiabatic Quantum Algorithms for Logistics Scheduling}

\author{Archismita Dalal}
\email{archismita.dalal@kipu-quantum.com}
\affiliation{Kipu Quantum, Greifswalderstrasse 226, 10405 Berlin, Germany}

\author{Iraitz Montalban}
\affiliation{Kipu Quantum, Greifswalderstrasse 226, 10405 Berlin, Germany}

\author{Narendra N. Hegade}
\affiliation{Kipu Quantum, Greifswalderstrasse 226, 10405 Berlin, Germany}

\author{Alejandro Gomez Cadavid}
\affiliation{Kipu Quantum, Greifswalderstrasse 226, 10405 Berlin, Germany}
\affiliation{Department of Physical Chemistry, University of the Basque Country UPV/EHU, Apartado 644, 48080 Bilbao, Spain}

\author{Enrique Solano}
\affiliation{Kipu Quantum, Greifswalderstrasse 226, 10405 Berlin, Germany}

\author{Abhishek Awasthi}
\email{abhishek.awasthi@basf.com}
\affiliation{BASF Digital Solutions GmbH, Pfalzgrafenstrasse 1, 67061 Ludwigshafen am Rhein, Germany}

\author{Davide Vodola}
\affiliation{BASF Digital Solutions GmbH, Pfalzgrafenstrasse 1, 67061 Ludwigshafen am Rhein, Germany}

\author{Caitlin Jones}
\affiliation{BASF Digital Solutions GmbH, Pfalzgrafenstrasse 1, 67061 Ludwigshafen am Rhein, Germany}

\author{Horst Weiss}
\affiliation{BASF SE, Pfalzgrafenstrasse 1, 67061 Ludwigshafen am Rhein, Germany}

\author{Gernot F\"uchsel}
\email{gernot.fuechsel@kipu-quantum.com}
\affiliation{Kipu Quantum, Greifswalderstrasse 226, 10405 Berlin, Germany}

\begin{abstract}
We study a job shop scheduling problem for an automatized robot in a high-throughput laboratory and a travelling salesperson problem with recently proposed digitized counterdiabatic quantum optimization (DCQO) algorithms. In DCQO, we find the solution of an optimization problem via an adiabatic quantum dynamics, which is accelerated with counterdiabatic protocols. Thereafter, we digitize the global unitary to encode it in a digital quantum computer. For the job-shop scheduling problem, we aim at finding the optimal schedule for a robot executing a number of tasks under specific constraints, such that the total execution time of the process is minimized. 
For the traveling salesperson problem, the goal is to find the path that covers all cities and is associated with the shortest traveling distance.
We consider both hybrid and pure versions of DCQO algorithms and benchmark the performance against digitized quantum annealing and the quantum approximate optimization algorithm (QAOA). 
In comparison to QAOA, the DCQO solution is improved by several orders of magnitude in success probability using the same number of two-qubit gates. Moreover, we implement our algorithms on cloud-based superconducting and trapped-ion quantum processors. Our results demonstrate that circuit compression using counterdiabatic protocols is amenable to current NISQ hardware and can solve logistics scheduling problems, where other digital quantum algorithms show insufficient performance.
\end{abstract}

\maketitle

\section{Introduction}

Quantum computing holds the promise of providing a significant advantage over classical computers for various industry-relevant problems, ranging from material simulation to optimization and machine learning \cite{bauer2020quantum}. In particular, the NP-hard class of optimization problems finds applications in finance, planning, and logistics \cite{HGL+23}. Despite the theoretical speedup claims of quantum search algorithms~\cite{Grover96}, no quantum algorithm has yet demonstrated an advantage over classical optimization algorithms using noisy intermediate-scale quantum (NISQ) computers. On the other hand, claims of ``quantum supremacy'' and ``quantum utility'' have been established using NISQ devices for non-industry-relevant problems \cite{AAB+19, ZWD+20, KEA+23}.
This gap in performance is mainly because the required depth of the quantum circuits for achieving an optimal solution exceeds the coherence time of a NISQ processor, and imperfect quantum gate operations further mars the quality of the solution. 
To counteract this drawback, one not only needs high-performing processors but also seeks an efficient problem-to-circuit encoding.
In this regard, we overcome the later challenge by using our circuit compression techniques. Specifically, we apply digitized counterdiabatic protocols to reduce the required circuit depth for tackling a combinatorial optimization problem on a gate-based quantum computer~\cite{hegade2021shortcuts, hegade2022digitized}, thereby demonstrating superior performance for various instances from industrial use cases.

The developing field of digitized counterdiabatic quantum computing finds applications in various problems ranging from combinatorial optimization, many-body ground state preparation to protein folding~\cite{hegade2021shortcuts,hegade2022digitized, chandarana2022protein,chandarana2022meta,chandarana2022digitized,hegade2023digitized, CMD+23, GZA+23, GDS+24}. This paradigm combines the advantages of counterdiabatic~(CD) protocols with the flexibility of digital quantum processors. CD protocols were first introduced in the analog quantum computing domain, where a CD Hamiltonian is added to the system's Hamiltonian to speed up the evolution without making any undesired transitions between the instantaneous eigenstates~\cite{demirplak2003adiabatic, berry2009transitionless, del2013shortcuts}. 
In the digital domain, the CD technique has been first proposed in Ref.~\cite{hegade2021shortcuts}, where it was used to prepare Bell and Greenberger-Horne-Zeilinger states on IBMQ hardware. 
Subsequently, this technique has 
been demonstrated to enhance the performance of digitized quantum annealing~(DQA)~\cite{hegade2022digitized} and quantum approximate optimization algorithm~(QAOA)~\cite{ chandarana2022digitized}.
Unlike their analog counterparts, digital quantum computers grant the ability to realize arbitrary CD terms. This flexibility paves the way for diverse Hamiltonians and superior control.
In this work we focus on digitized counterdiabatic quantum optimization~(DCQO)~\cite{hegade2022digitized} to solve logistics scheduling use cases.
By modifying the annealing Hamiltonian with additional CD terms, rapid state evolution can be achieved without succumbing to non-adiabatic transitions. This not only reduces the required circuit depth but also minimizes the impact of gate errors and decoherence.
Moreover, the hybrid version, namely hybrid-DCQO~(h-DCQO), integrates classical optimization techniques with DCQO to further enhance the solution quality and robustness~\cite{CMD+23}. 

We tackle two specific use cases from the NP-hard optimization problem class, where quantum computing is expected to offer a promising avenue for finding optimal or near-optimal solutions faster than their classical counterpart.
The first use case is an extension of a typical job-shop scheduling problem~(JSSP) to an industrial high-throughput laboratory process, and the second is the travelling salesperson problem~(TSP). 
For the JSSP use case considered in this work, the task is to determine the most efficient sequence of operations to be carried out by a robot, such that the different chemical samples are processed at different machines~\cite{LSJ+23}.
The complexity of these problems increases with the number of samples, machines used, and operational constraints, and finding an optimal solution using  classical algorithm requires an unfeasible growth in compute operations with problem size. 
Due to the large problem sizes of even the simplest non-trivial JSSP instances and the size limitations of gate-based NISQ devices, we restrict our study to a few `subproblems' of smaller sizes.
For TSP, where the task is to visit a specified number of cities in the least amount of time and return to the starting city, the required solution time also increases almost exponentially with the problem size, i.e.\ the number of cities involved~\cite{Lucas14}. Due to its more simplistic formulation, we could solve up to four-city TSP instances.

We employ DCQO and h-DCQO algorithms to obtain high-quality solutions for JSSP and TSP instances. 
By performing noiseless simulations, we demonstrate superior performance of DCQO and h-DCQO over DQA and QAOA, respectively, for all problem instances studied in this work. 
Notably, DCQO yields the exact solution atleast 3$\times$ faster than QAOA. 
Considering the variational training of QAOA and its multiple random initializations, DCQO is actually many orders of magnitude faster because of its pure quantum approach, whereby we need to run the quantum circuit just once.
We further decrease the depth of the DCQO circuits by restricting the evolution to the impulse regime, i.e.\ short evolution times, and discarding two-qubit gates with low-magnitude angles~\cite{CMD+23}. 
By adapting our original proposal for h-DCQO~\cite{CMD+23}, here we construct different variants of the ansatz to serve the problem instance better.
These techniques, along with efficient representation of the CD Hamiltonian in terms of hardware-dependent two-qubit gates, are essential towards successful implementations on NISQ hardware. 
We demonstrate the performance of DCQO and h-DCQO circuits using one 16-qubit JSSP instance, and 9-qubit and 16-qubit TSP instances. IonQ's trapped-ion hardware successfully solves all three instances, whereas IBMQ's superconducting circuit could only solve the 9-qubit instance. Thus, our results demonstrate that counterdiabatic techniques are better able to utilize NISQ hardware for small-scale optimization problems.

The work is organized as follows.
We provide a brief background on the two use cases that we tackle in this work, namely, JSSP and TSP, in \S\ref{sec:usecases}.
Next in \S\ref{sec:methods}, we explain digitized counterdiabatic optimization algorithms, both pure quantum and hybrid quantum-classical versions, and the circuit-depth reduction techniques that we use for successfully solving optimization problems on NISQ hardware.
The solutions obtained from using (h-)DCQO for JSSP and TSP are discussed in \S\ref{subsec:jssp_results} and \S\ref{subsec:tsp_results}, respectively, along with their comparison against the solutions obtained from existing NISQ optimization algorithms.
Finally, we conclude with a future scope and outlook in \S\ref{sec:conclusion}. 

\section{Problem descriptions}
\label{sec:usecases}
In this section, we describe the two use cases in logistics scheduling that we tackle in this work. First we discuss the scheduling problem in a high-throughput laboratory process, and the challenges associated with solving its relevant instances on NISQ hardware. Next we provide a brief background on TSP and its existing quantum solutions.
\subsection{Job-shop scheduling problem}
\begin{figure}
    \begin{center}
    \includegraphics[width=0.9\linewidth]{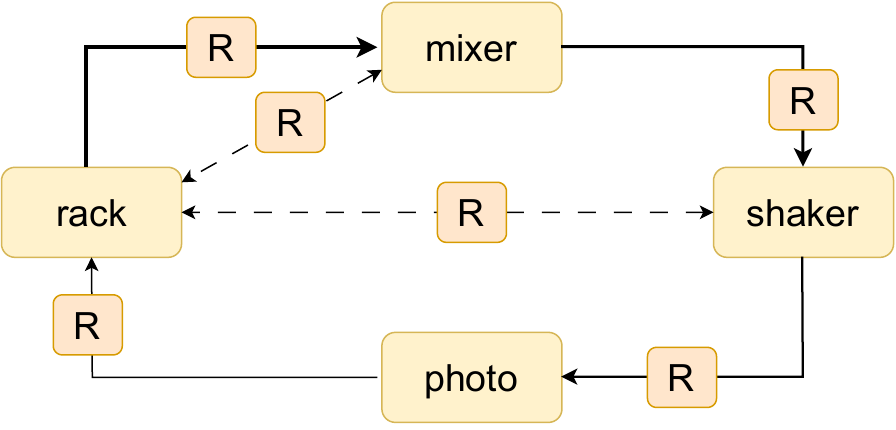}
    \end{center}
    \caption{A simplified job-shop scheduling problem. The robot~R carries a chemical sample between the four stations (rack, mixer, shaker and photo booth) for processing in a pre-defined order, i.e.\ rack $\rightarrow$ mixer $\rightarrow$ shaker $\rightarrow$ photo booth. The path is shown by solid lines with decreasing thickness from the first to the last segment. Moreover, the samples can be transported back to the rack from mixer and shaker for storing in between the processing, which is shown by dashed lines.
   }
    \label{fig:jssp_setup}
\end{figure}
The JSSP describes here an automatized process in a chemical laboratory and is schematically represented in Fig.~\ref{fig:jssp_setup}. The goal is to find an optimal sequence of autonomous operations that minimizes the total time required to process a given number of chemical samples~($n_\text{s}$) in a laboratory~\cite{LSJ+23}. The setup here consists of four stations, namely rack, mixer, shaker and photo booth, and a single robot is tasked with moving samples between the stations one at a time; see Fig.~\ref{fig:jssp_setup}. Each sample needs to be processed in a particular order: mixed, shaken and then photographed multiple times in pre-defined intervals. The rack stores all the samples in the beginning and at the end of its processing, as well as any time in between if needed. There are other constraints including pre-determined processing times at mixer and shaker, and required time-gap between two successive photographs for each sample. Thus, the goal is to find a sequence of instructions for the robot that minimizes the sum of processing times of each sample under multiple constraints.

In the simplest non-trivial setting, one can consider two samples, i.e.\ $n_\text{s}=2$, one photograph for each sample and a time horizon~($n_\text{t}$) of 30 units. The total number of machines~($n_\text{m}$) used is 3, which are the mixer, the shaker and one photo station. Based on a particular problem formulation~\cite{LSJ+23}, the number of binary variables required is $2n_\text{m}n_\text{s}n_\text{t}$.
Thus, the simplest problem setting results in a quadratic unconstrained binary optimization~(QUBO) problem instance of size 360, out of which many variables can be eliminated by imposing the problem constraints. But a quantum optimization algorithm would still require hundreds of qubits to solve this JSSP instance, which is beyond what current NISQ devices are capable of and an adaption of the problem size needs to be carried out.  

In our work, we aim to test the DCQO and h-DCQO algorithms not only on simulators, but also on available digital hardware.
Therefore, we decompose the full QUBO matrix into sub-QUBOs of size 16, which are implementable on available NISQ devices. These sub-QUBOs are selected in a manner that the connectivity between the pairs of variables is very high, which requires a well connected hardware to embed the problem Hamiltonian. After solving all these subproblems, the solution of the full QUBO problem is updated and the decomposition-composition process is repeated until some convergence criteria is met. This iterative technique is commonly known as the large neighborhood search in classical optimization~\cite{PR19}.
In the quantum domain, D-Wave has long borrowed this idea to create their quantum-classical Hybrid Solver~\cite{RVWH16,BRR17}, and more recently QAOA has been implemented in a similar setting~\cite{PHL+23}.
In this work, we focus on solving ten of the subproblems from a JSSP instance on gate-based quantum devices, and analyze the performance of the studied quantum algorithms in achieving good results for densely connected QUBOs.
We present simulation results for all these ten 16-qubit instances to compare DCQO and DQA. 
Additionally, we compare h-DCQO against QAOA for those ten instances.
For one of the instances, we provide cloud-based demonstration results from IonQ hardware.

\subsection{Travelling salesperson problem}
\begin{figure}
    \centering
    \includegraphics[width=0.8\linewidth]{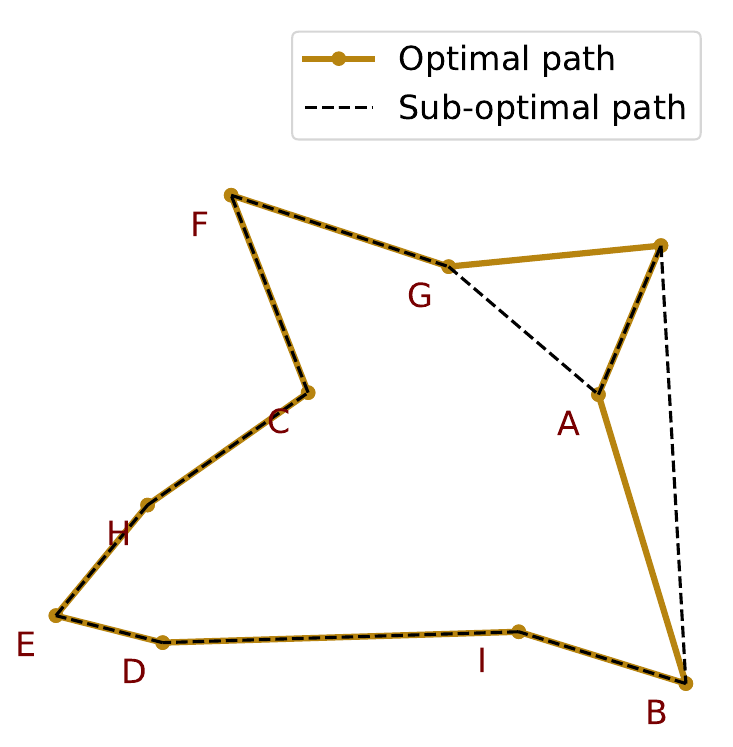}
    \caption{A TSP instance with 10 cities A to J. The shortest round trip is shown using solid (gold) thick lines. An alternative round trip with greater distance is shown with dashed (black) thin lines.
}
    \label{fig:tsp_germany}
\end{figure}

TSP is a prototypical combinatorial optimization problem falling into the complexity class of NP-hard~\cite{OYD20}. The task is to find the shortest round trip that a salesperson can take to cover a given number of cities; see Fig.~\ref{fig:tsp_germany}. Besides planning and logistics, TSP finds applications in manufacturing and DNA sequencing \cite{10051206, JOUR}. The scaling of classical-compute time is almost exponential with the number of cities, making TSP an ideal candidate for quantum computing. Although Grover's quantum algorithm promises a quadratic speedup for search~\cite{Grover96}, this claim is far beyond reach due to noise and limited size of current quantum hardware. Thus, we employ DCQO and h-DCQO to solve TSP instances requiring up to 16 qubits in this work.

We follow the one-hot encoding technique to map TSP into a QUBO problem~\cite{Lucas14}. As per this encoding, $n^2$ bits are required to represent a $n$-city TSP. In a bitstring $\bf{x}$ of length $n^2$, $x_{t,c}=1 $ if the salesperson visits city~$c$ in the $t$ step of their path. For example, with $n=4$ cities, a path $0\to2\to1\to3$ is represented by $1000\ 0010\ 0100\ 0001$. With a trivial mapping of one bit to one qubit, we thus end up using $N_\text{q} = n^2$ qubits. Due to such a big overhead in the number of qubits, and consequently possible outcomes, most ($2^{N_\text{q}}-\frac{n!}{2}$) bitstrings are infeasible outcomes for TSP. This makes finding the shortest path  even more challenging. Utilizing quantum algorithms for solving simplistic 3-city and 4-city TSPs not only helps us in verifying whether constraints and objectives can be converged with these algorithms, but also makes implementation of TSP on NISQ hardware feasible.

Only a few quantum approaches have been mentioned in literature to solve TSP instances. In the analog quantum computing regime, D-Wave's quantum annealing provides a straightforward approach, but can only solve up to 8-city TSP instances due to device noise and sparse qubit connectivity~\cite{Jain21}.
In the digital regime, quantum phase estimation~\cite{SSB+18} and QAOA~\cite{KCPW21} have been used to address these problems.
A recent numerical study demonstrates the advantage of using a superior encoding, i.e.\ $n\log n$ instead of $n^2$, for solving 4-city TSP instances using noiseless simulations of QAOA~\cite{RSS+24}. Additionally, using a different variant of QAOA circuit has shown superior performance in noiseless simulation of TSP instances with up to 16 qubits~\cite{QBE+23}. 
As the $n\log n$ encoding requires a lot more two-qubit gates~\cite{RSS+24}, we restrict the scope of this work to the trivial $n^2$ encoding. 

\section{Methods}
\label{sec:methods}
In this section, we elaborate on the design and analysis of counterdiabatic algorithms, both the pure quantum implementation and the quantum-classical variational implementation, for solving combinatorial optimization problems. Additionally, we discuss methods for further reducing DCQO circuit depths and making them amenable to NISQ hardware. In particular, we discuss the concepts of impulse regime and gate-angle thresholding for DCQO circuit construction, and explain how the Pauli operations in the CD Hamiltonian can be efficiently implemented on IBMQ and IonQ hardware. 

A real-variable quadratic optimization problem can be approximately converted into a binary-variable quadratic optimization problem using real-to-binary encoding. Furthermore, any constraint can be added to the optimization cost function by appropriate Lagrange multiplier. This leads to a QUBO problem, where the goal is to obtain the bit-string~$x_0$ that minimizes or maximizes the objective function~\cite{GKD18}
\begin{equation}
    f(x) = x^T Q x, 
\end{equation}
for a problem-dependent real-valued symmetric matrix~$Q$. This QUBO problem is equivalent to finding the ground state of an Ising spin-glass Hamiltonian~\cite{BCMR10}
\begin{equation}
\label{eq:Hprob}
    H_\text{p} = \sum_{i<j} J_{ij} \sigma_z^i \sigma_z^j + \sum_i h_i \sigma_z^i, 
\end{equation}
where $\sigma_z^i$ denotes the Pauli Z matrix acting on the $i$-th spin, and the coefficients $J_{ij}=\frac{1}{2}Q_{ij}$ and $h_i=-\frac{1}{2}Q_{ii} -\sum_j\frac{1}{2}Q_{ij}$. Problems encoded in Ising Hamiltonians are suitable for a quantum computing treatment and several quantum optimization algorithms have been developed including adiabatic quantum optimization~\cite{FGG+01}, quantum annealing~\cite{KN98}, variational quantum eigensolver~\cite{MBB+18} and QAOA~\cite{FGG14}, which are briefly discussed in Appendix~\ref{appendix:qo}.

\subsection{Digitized counterdiabatic quantum optimization}
\label{subsec:dcqo}
A rapid driving of the system Hamiltonian $H_\text{ad}(t)$~\eqref{eq:Hanneal} in adiabatic quantum optimization results in unwanted transitions between instantaneous eigenstates, and, consequently, a sub-optimal solution of the optimization problem. 
In this regard, counterdiabatic driving is effected by adding a CD Hamiltonian as
\begin{equation}
\label{eq:counterdiabatic}
    H(t) =
    H_\text{ad}(t) +  
    \dot{\lambda}(t) A_{\lambda}(t),
\end{equation}
where the adiabatic gauge potential~$A_{\lambda}(t)$ is responsible for suppressing the non-adiabatic excitation during the Hamiltonian evolution and $\dot{\lambda}(t)$ is the time-derivative of the scheduling function governing that evolution.
As the exact computation of~$A_{\lambda}(t)$ is computationally expensive
\cite{berry2009transitionless}, approximate techniques for constructing and realizing counterdiabaticity have been proposed~\cite{sels2017minimizing, claeys2019floquet, hatomura2021controlling, takahashi2023shortcuts}.
In particular, one can approximate the gauge potential using a series of nested commutators (NC) as
\begin{equation}
\label{eq:commutator}
    A_{\lambda}^{(l)}(t) = i\sum_{k=1}^l \alpha_k(t) \underbrace{[ H_\text{ad}, [ H_\text{ad},\dots [ H_\text{ad},}_{2k-1} \partial_\lambda H_\text{ad} ] ] ]\;,
\end{equation}
where $l$ sets the maximum order of approximation and $\alpha_k(t)$ is the CD coefficient corresponding to the $k$th-order approximation.

A high-order ($l\ge2$) NC expansion of the gauge potential~\eqref{eq:commutator} leads to a Hamiltonian with many-body (>2) Pauli terms, whose implementation with only one- and two-qubit gates dramatically increases the circuit depth. 
Thus we restrict to only first-order NC expansion~\cite{hegade2022digitized}, where
\begin{align}
    A_\lambda^{(1)}(t) &= i\alpha_1(t) \left[H_\text{i}, H_\text{p}\right] \nonumber\\
    &=- 2 \alpha_1(t) \left[ \sum_i h_i \sigma_y^i + \sum_{i<j} J_{i j}\left(\sigma_y^i \sigma_z^j+\sigma_z^i \sigma_y^j\right) \right]
    \label{eq:gauge_1}
\end{align}
for an initial transverse field Hamiltonian~\eqref{eq:Hinit}.
With this lowest order expansion, we can analytically express the CD coefficient as
\begin{align}
\alpha_1(t) = -\frac{1}{4} \frac{ \sum_i h_i^2+ \sum_{i<j} J_{i j}^2 }{\gamma(t)} \, ,
\end{align}
where
\begin{equation}
\begin{split}
    \gamma(t) =& (1-\lambda(t))^2\left(\sum_i h_i^2+4 \sum_{i \neq j} J_{i j}^2\right) \\ 
    &+ \lambda^2(t)\bigg[\sum_i h_i^4+\sum_{i \neq j} J_{i j}^4+6 \sum_{i \neq j} h_i^2 J_{i j}^2\\ 
    &+6 \sum_{i<j<k}\left(J_{i j}^2 J_{i k}^2+J_{i j}^2 J_{j k}^2+J_{i k}^2 J_{j k}^2\right)\bigg].
\end{split}
\end{equation}
The higher-order NC expansion terms can be numerically evaluated using \texttt{qiskit}'s functionalities.
These approximations to $A_\lambda$ does not guarantee full suppression of non-adiabatic transitions during the evolution, but reduces their probability to an extend that allows achieving good results on NISQ hardware for combinatorial optimization problems including factorization~\cite{hegade2023digitized}, portfolio optimization~\cite{CMD+23} and p-spin models~\cite{GZA+23}.

To implement counterdiabatic driving on a digital or gate-based quantum computer, we discretize the total evolution time~$T$ of the full Hamiltonian~\eqref{eq:counterdiabatic} into $N$ Trotter steps, with each step being applied for a duration $\Delta t=\frac{T}{N}$.
Furthermore, with a short-$T$ approximation, \ $|\lambda(t)| \ll |\dot{\lambda}(t)|$ for most of the evolution, and consequently $H(t) \approx \dot{\lambda}(t) A_{\lambda}(t)$~\cite{KOLODRUBETZ20171}.
Using Eq.~\eqref{eq:gauge_1} and $\lambda(t)=\sin^2\left(\frac{\pi}{2}\sin^2\left(\frac{\pi t}{2T}\right)\right)$, DCQO is effected by the unitary operator
\begin{eqnarray}
    U_\text{QO} &\approx& 
    \prod_{m=1}^N \exp\left(-i\frac{\pi \sin(\pi \frac{m}{N}) \sin(\pi\sin^2(\pi \frac{m}{2N}))}{2N}A_\lambda^{(1)}(m\Delta t)\right),
\label{eq:dig_evo}
\end{eqnarray}
which is then implemented using one- and two-qubit gates.
Our chosen scheduling function not only ensures the quantum annealing protocol by ramping up from 0 to 1 within $T$, both first and second derivatives of $\lambda(t)$ also vanish at the boundary points~\cite{claeys2019floquet}.
The above equation shows that the evolution remains independent of~$T$ if $N$ is fixed; this feature is independent of the expansion order of the gauge potential~\eqref{eq:commutator}.
For an ideal simulation of DCQO, the success probability depends on $N$, $T$ (or $\Delta t$) and the number of qubits~$N_\text{q}$. As previously demonstrated, the success probability drops exponentially with increasing~$N_\text{q}$, but with a polynomial speedup as compared to DQA~\cite{hegade2022digitized}. 
In practice, we use a few Trotter steps and different circuit-reduction techniques~(\S\ref{subsec:red}) to keep the DCQO circuit depth within the coherence limit of current quantum processors.

\subsection{Hybrid-DCQO}
\label{subsec:hdcqo}
\begin{figure*}
\centering
  \includegraphics[width=.75\linewidth]{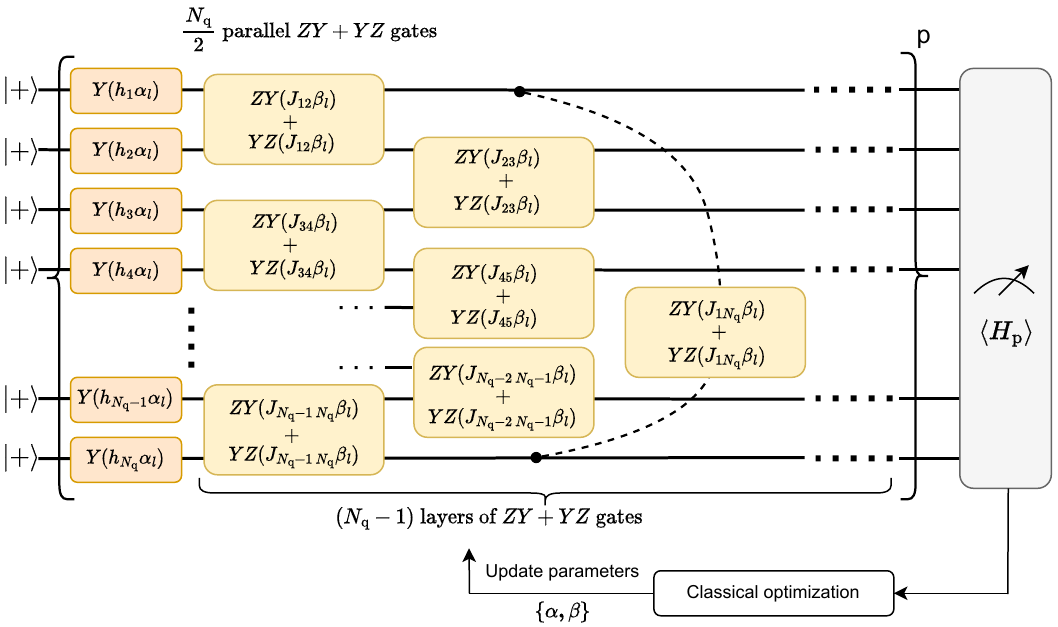}
  \caption{
  A $p$-layer hybrid-DCQO circuit assigns two parameters per layer~$l$, namely~$\alpha_l$ and $\beta_l$ for one-body and two-body terms in the CD Hamiltonian, respectively. The light orange block represents the single-qubit evolution~$\exp(-ih_i\alpha_l\sigma_y^i)$.
  The light yellow block represents the combined evolution~$\exp(-i\beta_l J_{ij}\sigma^i_z\sigma^j_y)\exp(-i\beta_l J_{ij}\sigma^i_y\sigma^j_z)$ on two qubits~$i$ and $j$.
  The circuit is initialized by an uniform superposition of the computational basis.
  The parameters are updated in an iterative manner based on the classical optimization of the cost function $\langle{\mathbf{\alpha},\mathbf{\beta}}|H_\text{p}|{\mathbf{\alpha},\mathbf{\beta}}\rangle$. 
  }
  \label{fig:hdcqc_ansatz}
\end{figure*}
The classical-quantum hybrid version of the DCQO algorithm is called hybrid-DCQO~(h-DCQO), which is a variational quantum algorithm based on a CD-inspired ansatz selection~\cite{chandarana2022protein}. The development of this algorithm is motivated by the observation that the success probability of the pure approach, i.e.\ DCQO, exponentially decreases with increasing number of qubits.
In this algorithm, the time-dependent coefficients, i.e.\ $2\dot\lambda(t)\alpha_1(t)$~\eqref{eq:gauge_1}, in the CD Hamiltonian are taken to be time-independent free parameters that can be variationally trained, as opposed to explicitly calculating them as done in the pure quantum approach. In an earlier version of the hybrid algorithm, the coefficients of the full Hamiltonian~\eqref{eq:counterdiabatic} were parameterized, leading to digitized-counterdiabatic-QAOA~(DC-QAOA) that numerically demonstrates superior performance over QAOA for random Ising spin models, MaxCut problem and p-spin models~\cite{chandarana2022digitized}.
In comparison to DC-QAOA, the low-depth ansatz in the h-DCQO algorithm is composed of only the CD term, making it amenable to current quantum hardware. In particular, the h-DCQO algorithm has outperformed state-of-the-art algorithms for protein folding on both trapped ions and superconducting circuits~\cite{chandarana2022protein}, for portfolio optimization on trapped ions~\cite{CMD+23} and for p-spin model on superconducting circuits~\cite{GZA+23}.

We now briefly describe the h-DCQO algorithm; see Fig.~\ref{fig:hdcqc_ansatz}.
A simple h-DCQO ansatz is parameterized by $\mathbb{\alpha}$ and $\mathbb{\beta}$, which correspond to the one- and two-body Hamiltonian terms, respectively, of the first-order NC expansion~\eqref{eq:gauge_1}. These parameters are initialized either with random values or with the analytically calculated values~\eqref{eq:gauge_1}, which we refer to as ``warm-starting'' h-DCQO. Then we variationally update the parameters such that the expectation value of the problem Hamiltonian~\eqref{eq:Hprob} is minimized. 
Although a two-parameter h-DCQO ansatz ensures equal number of parameters for both QAOA and h-DCQO ansatze, the Molmer-Sorensen~(MS) gate count for one layer of h-DCQO is twice than that of QAOA. 
In spite of this increased gate count per layer, our h-DCQO algorithm outperforms QAOA in portfolio optimization problems by reducing the required number of layers for achieving a target performance~\cite{CMD+23}. 
On the other hand, the CX gate count per layer for QAOA and h-DCQO can be kept equal with efficient decomposition strategy~(\S\ref{subsec:red}).
In addition to the two-parameter h-DCQO ansatz, we also use an ansatz with $(N_\text{q}+1)$-parameters, where all one-body terms are independently parameterized.

\subsection{Circuit-depth reduction techniques}
\label{subsec:red}
\begin{figure}
\centering
    \includegraphics[width=.95\linewidth]{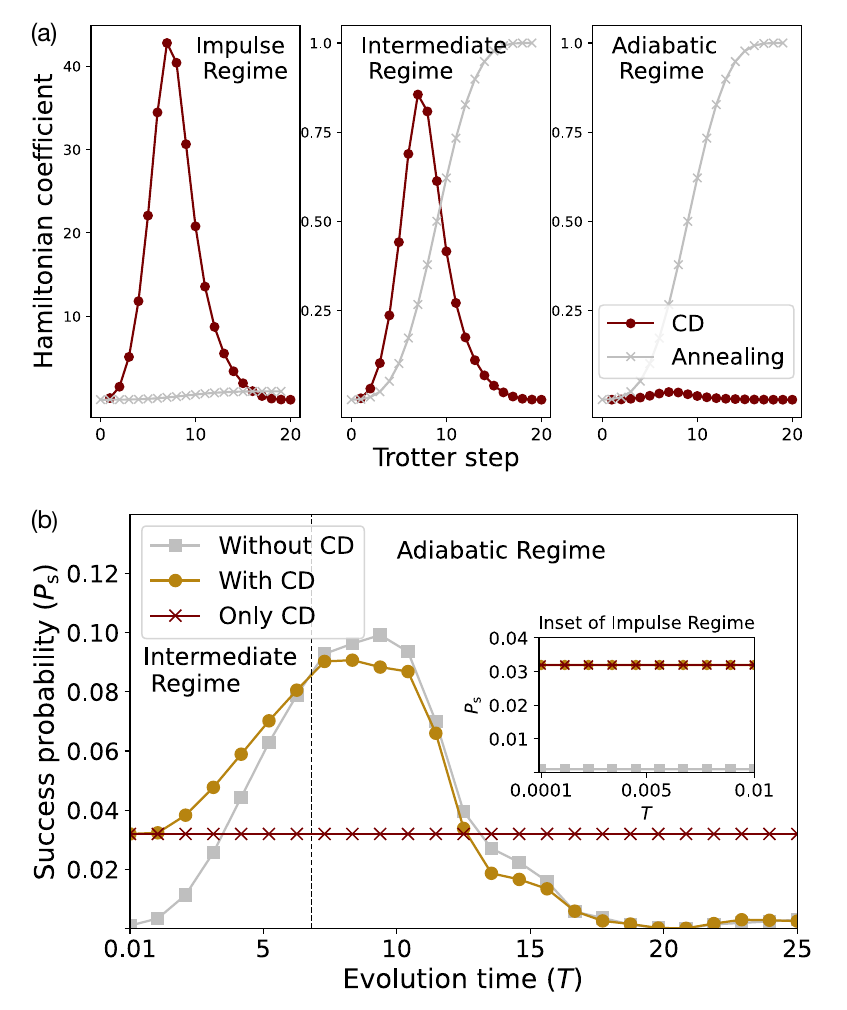}
    
      \caption{
      Impulse regime analysis for a random 10-qubit Ising spin-glass model.
      (a) Hamiltonian coefficients vs. Trotter step for the three operational regimes. $T=0.005,~0.25,~10$ for impulse, intermediate and adiabatic regimes, respectively, and $N=20$. Notice the different y-axis scales of the three plots.
      (b) Dependence of success probability on evolution time~$T$, with $N = 20$, for three different Hamiltonian evolutions. 
      Without CD, With CD and Only CD corresponds to evolutions of the annealing Hamiltonian, full Hamiltonian and CD Hamiltonian, respectively.
      The inset magnifies the impulse regime where the performance of Only CD Hamiltonian matches that of the total Hamiltonian. 
      }
  \label{fig:cdcontrib}
\end{figure}
As mentioned in \S\ref{subsec:dcqo}, we consider DCQO in the impulse regime, and
now we demonstrate the validity of this approximation using DCQO simulations for a 10-qubit random Ising spin-glass problem.
The evolution of the full Hamiltonian~\eqref{eq:counterdiabatic} can be broadly categorized into three regimes, namely impulse, intermediate, and adiabatic regimes, based on the evolution time~$T$~\cite{CMD+23}.
From Fig.~\ref{fig:cdcontrib}(a), we observe that the maximum coefficient of the CD Hamiltonian is about $40\times$ than that of the annealing Hamiltonian~\eqref{eq:Hanneal}, i.e.\ $\lambda(t)$, in the impulse regime; whereas, the ratio reverses in the limit of adiabatic evolution. Thus the importance of the CD Hamiltonian in comparison to the annealing Hamiltonian indeed increases with decreasing the evolution-time parameter from $T=10$ to $T=0.005$. 
This is also evident from Fig.~\ref{fig:cdcontrib}(b), where we not only show the three operational regimes, but also highlight that CD-assisted evolution achieves much  higher success probabilities at very short $T$. 
As expected from Eq.~\eqref{eq:dig_evo}, the success probability obtained from Only CD evolution is independent of $T$ for a fixed $N$.
Whereas, the success probabilities obtained from both annealing and full Hamiltonians
increase with increasing $T$ up to some maximum value, and then the probabilities start dropping due to increasing Trotter error. 

\begin{figure}[t!]
\centering
\includegraphics[width=.9\linewidth]{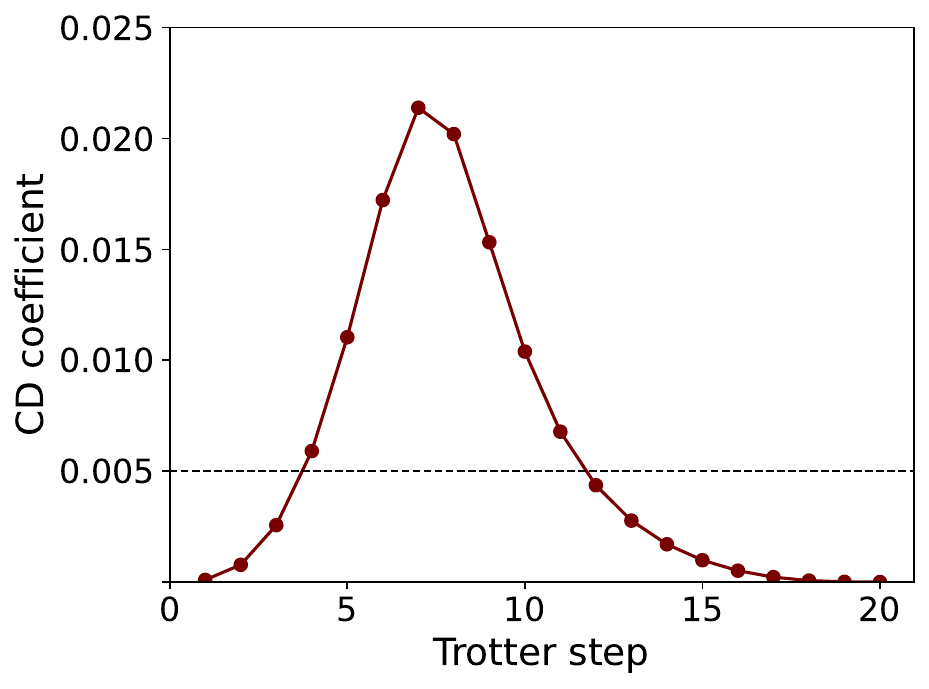}
      \caption{
      Dependence of CD coefficient on Trotter step, with the dashed line separating significant from insignificant contribution.
      }
  \label{fig:cdcontrib1}
\end{figure}

We can further eliminate a few unitaries from the trotterized Only CD evolution based on their corresponding Hamiltonian coefficients.
In~Fig.~\ref{fig:cdcontrib1}, we show that the contribution of the initial and final Trotter steps of the CD Hamiltonian can be neglected as compared to the middle steps. Thus by setting a threshold, say 0.005 as marked in~Fig.~\ref{fig:cdcontrib1}, we can shorten a DCQO circuit to effectively half the required depth.
Additionally, if the angle associated with a single- or two-qubit gate is smaller than a chosen threshold, we omit the corresponding gate from the DCQO circuit and thus further shorten its depth.
This approximation is valid because the DCQO algorithm would then only apply the gates required to modify the state of the system up to a certain resolution set by the threshold. The smallest angle that IonQ hardware can accurately identify is 0.00628~\cite{native_gates}, whereas for IBMQ, the value is not publicly available. Thus we choose a gate-cutoff threshold of 0.1  this work.

We can additionally compress the depth of the DCQO circuit by using hardware-dependent transpilation techniques.
Although the evolution of each two-body term, i.e.\ $YZ$ and $ZY$, requires two CX gates, we can efficiently implement their combined operation on IBMQ hardware using only two instead of four CX gates. 
This decomposition can be achieved by first constructing a circuit where the rotation operators~$R_{YZ}$ and $R_{ZY}$ are placed consecutively for each pair of interacting qubits, and then using the generic \texttt{qiskit} transpiler.
On a circuit level there are implications of reordering Pauli terms of a Hamiltonian, which by default are ordered lexicographically in \texttt{qiskit}, due to the error introduced from finite number of Trotter steps during the digitization process~\cite{TLM+19}. 
For IonQ implementation, we make use of its native two-qubit MS gate. The evolution of each of the two-body Hamiltonian terms can be expressed with only one MS gate. In particular, the entangling  gate~$R_{YY}(\theta)$ can be implemented using MS gate as
\begin{widetext}
\begin{equation}
R_{YY}(\theta) =  
\begin{cases}\!
    MS(\pi/2,\pi/2, \theta) & \theta \leq \frac{\pi}{2}\\
(GPi(\pi/2) \otimes  GPi(\pi/2))\times MS(3\pi/2,\pi/2, \pi-\theta), & \frac{\pi}{2} < \theta \leq \pi \\
(GPi(\pi/2) \otimes  GPi(\pi/2))\times MS(\pi/2,\pi/2, \theta-\pi), & \pi < \theta \leq \frac{3\pi}{4} \\
MS(3\pi/2,\pi/2, 2\pi - \theta), &\theta > \frac{3\pi}{2},
\end{cases}
\end{equation}
\end{widetext}
where $GPi$ and $GPi2$ are native single-qubit gates.
The above entangling operation can then in turn be used to construct the relevant operations 
\begin{align}
    R_{ZY}(\theta) &= (GPi2(\pi) \otimes  I)R_{YY}(\theta)(GPi2(0) \otimes  I),\nonumber\\
    R_{YZ}(\theta) &= (I \otimes  GPi2(\pi))R_{YY}(\theta)(I \otimes  GPi2(0))
\end{align}
for the DCQO circuit. Thus, we can represent the combined evolution of $YZ$ and $ZY$ using only two native two-qubit gates, on both superconducting circuits and trapped ion hardware.

\section{Results}
\label{sec:results}
In this section, we discuss quantum solutions to the logistics scheduling use cases introduced in \S\ref{sec:usecases}, namely job-shop scheduling and travelling salesperson problems. 
We apply both pure quantum and hybrid quantum-classical counterdiabatic algorithms to solve different instances of these use cases, and additionally compare those solutions against the ones obtained using DQA and QAOA. 
Finally, we report hardware implementation results obtained from two different gate-based quantum processors, namely IonQ's trapped ions and IBMQ's superconducting circuits, available through cloud services.
The device calibration data at the time of our demonstration can be found in Appendix \ref{appendix:cali}.
In this work, we use the typical performance metrics, namely success probability (SP) and approximation ratio (AR), for quantum optimization algorithms because we can compute the exact solutions of the tackled problem sizes. 
SP is the probability of finding the ground-state bitstrings, i.e.\ the bitstrings corresponding to the exact solutions, from all possible bitstrings in the output distribution. Whereas, AR measures the fraction of near-minimal bitstrings, and it is calculated as ratio between the average energy of the output distribution and the ground-state energy.

\subsection{(h-)DCQO for JSSP}
\label{subsec:jssp_results}
\begin{figure}
    \centering
    \includegraphics[width=0.95\linewidth]{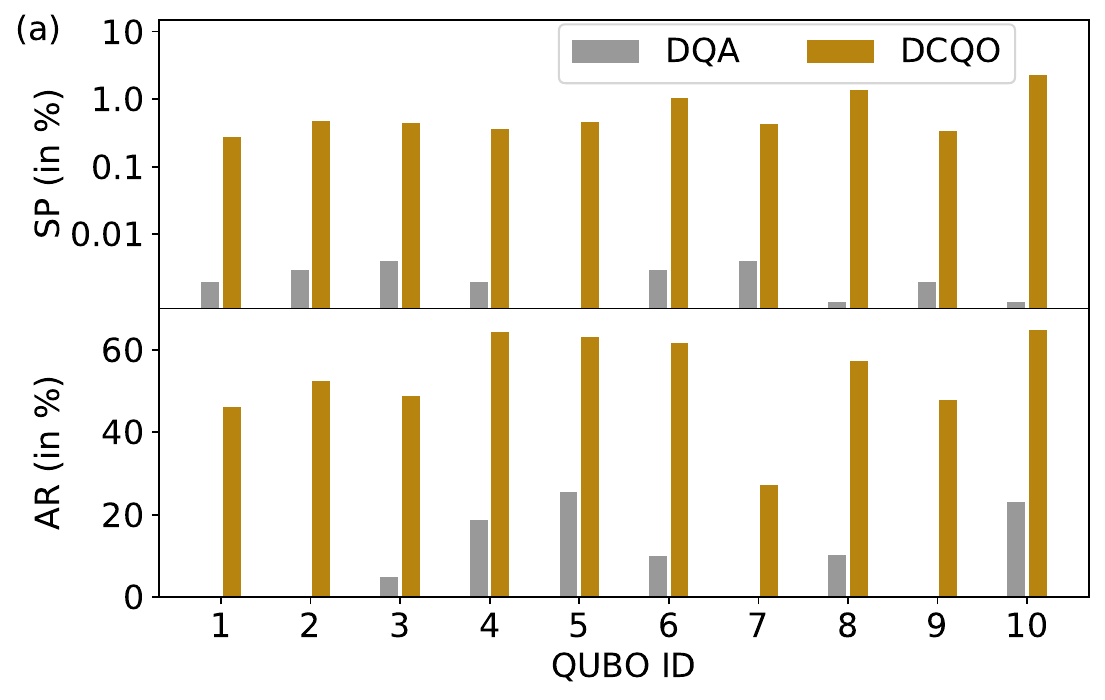}\\
    \includegraphics[width=0.95\linewidth]{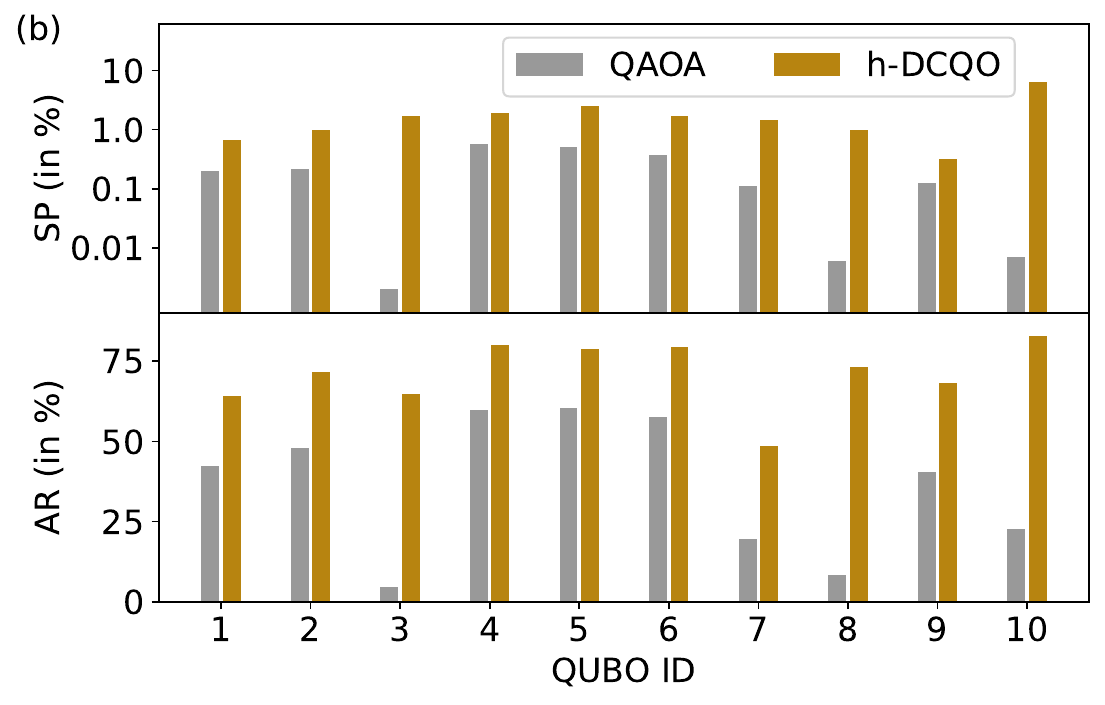}
    \caption{Ideal simulation results comparing counterdiabatic algorithms against existing NISQ algorithms for solving 10 subproblems of JSSP. (a) DCQO circuits with $T=0.1$ and $N=4$ has the same number ($\approx 1440$) of CX gates as DQA circuits with $T=1.0$ and $N=6$.
    (b) With one layer, each of the h-DCQO and QAOA circuits has effectively 240 CX gates. }
\label{fig:compare_jssp}
\end{figure}
Due to the intractability of the full job-shop scheduling problem~(\S\ref{sec:usecases}~A), we address only ten subproblems of size 16. These QUBO instances are fully-connected, making them challending to be solved using NISQ hardware. 
For these instances, we first compare DCQO~(h-DCQO) against DQA~(QAOA) using noiseless simulations.
In Fig.~\ref{fig:compare_jssp}(a), we observe that with an equal circuit depth, DCQO has a SP (AR) of at least 100$\times$ ($2\times$) greater than that of DQA for all the ten instances. 
In particular, for QUBO ID 5, DQA fails to find the ground state.
Due to such extremely low SPs and high CX counts, DQA is expected to fail in achieving the optimal solutions on NISQ hardware for these 16-qubit instances. 
Moreover for QUBO instances 1, 2, 7 and 9, DQA yields negatives ARs, implying extremely poor solutions.
On the other hand, even after omitting about 80\% CX gates using the gate-cutoff technique~(\S\ref{subsec:red}), DCQO is still able to find the ground states with SP exceeding 0.1\% in noiseless simulations. 
In the following paragraph we show that DCQO, with gate cutoff, also performs well on NISQ hardware.
In Fig~\ref{fig:compare_jssp}(b), we observe that warm-started h-DCQO~(\S\ref{subsec:hdcqo}) always yields superior results, in terms of both SP and AR, than one run of randomly-initialized QAOA. Although one can perform multiple random initialization of QAOA to search for a better result, we empirically observe that QAOA always falls behind h-DCQO. 

\begin{figure}[]
    \centering
    \includegraphics[width=\linewidth]{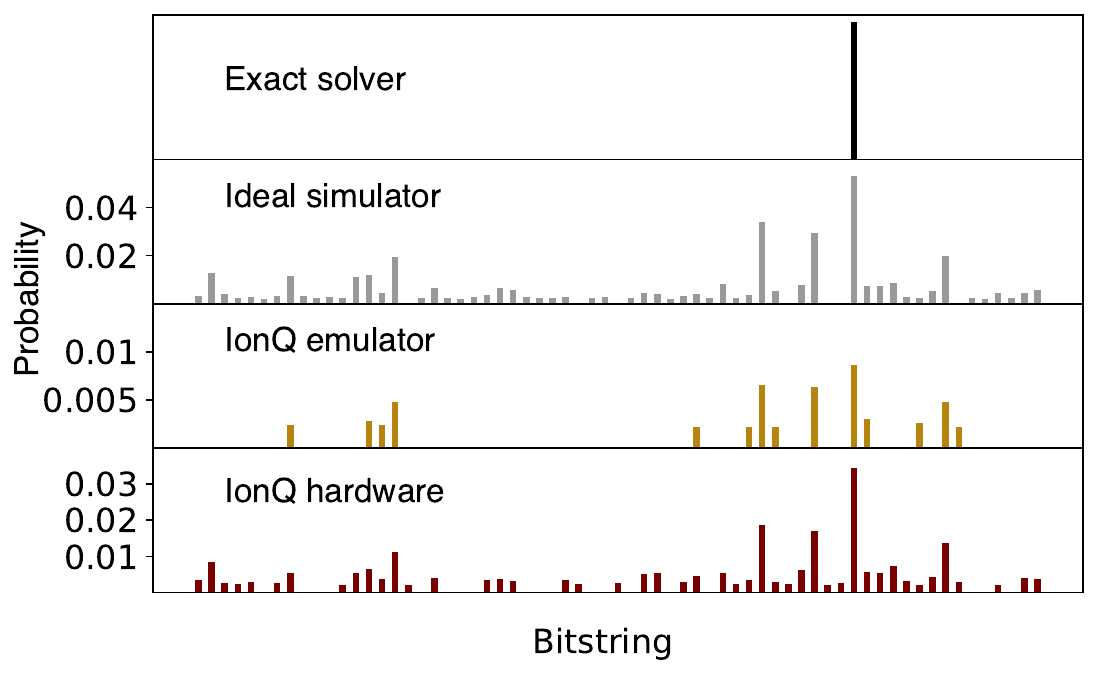}
    \caption{DCQO results for a 16-qubit subproblem of JSSP. The plot shows only the top few probabilities from the whole probability distribution. For $T=0.002$, $N=2$ and a gate-cutoff threshold of 0.1~(\S\ref{subsec:red}), the DCQO circuit has 142 MS gates. The IonQ emulator uses the noise model of \texttt{Aria-2}, which is also the hardware used for our demonstration. 
    }
    \label{fig:dcqo_jssp}
\end{figure}
Among the ten subproblems, we notice that the last instance is easy to solve in a sense that it has the highest SP among the ten instances studied. We further test the performance of DCQO for this instance both on a noisy simulator and a real quantum hardware to understand the impact of noise on DCQO's performance and gauge the usability of DCQO for a 16-qubit optimization problem. 
In Fig.~\ref{fig:dcqo_jssp}, we notice that DCQO can identify the true ground state in a noisy simulation, although the success probability is reduced to one-quarter as compared to the noiseless case. Using a total of 5000 measurements and noise mitigation, particularly debiasing, IonQ's \texttt{Aria-2} yields the exact solution of the QUBO problem with 3.5\% probability. This performance is possible due to the all-to-all qubit connectivity of the hardware, along with its high two-qubit gate fidelity and coherence time.
In contrast, the sparse qubit connectivity of \texttt{ibm\_brisbane} requires twice as much two-qubit gates in the transpiled circuit. This massive increase in the two-qubit gate count, along with lower gate fidelities of IBMQ's hardware, leads to very low quality in the solution for this problem instance.

\begin{figure}
    \centering
    \includegraphics[width=\linewidth]{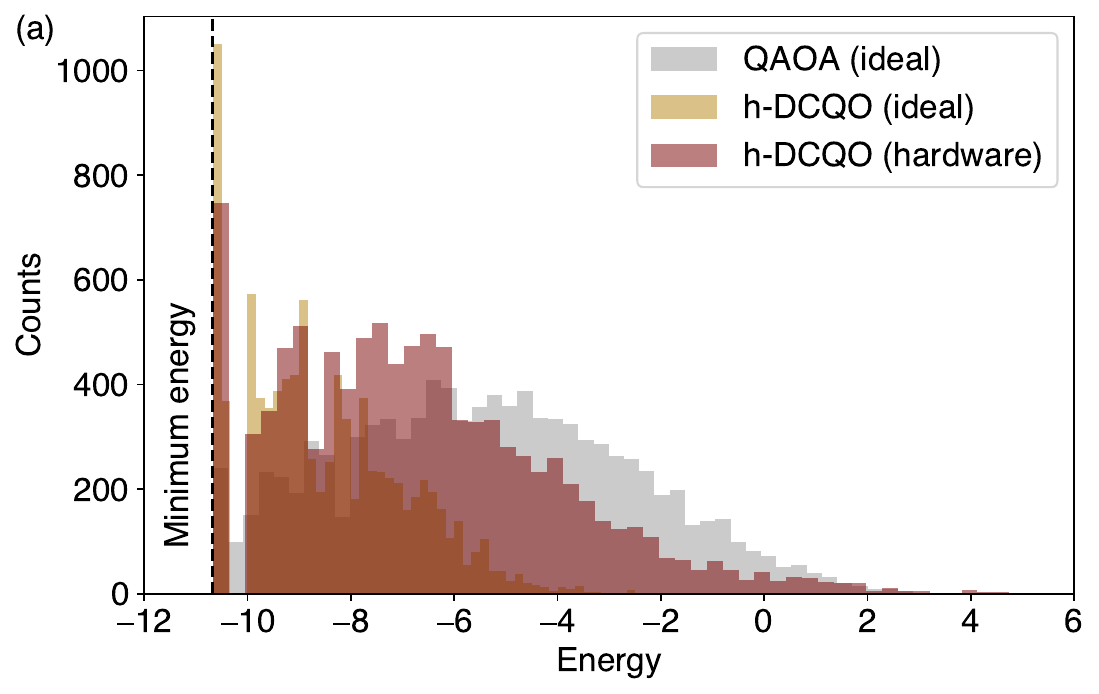}\\
    \includegraphics[width=\linewidth]{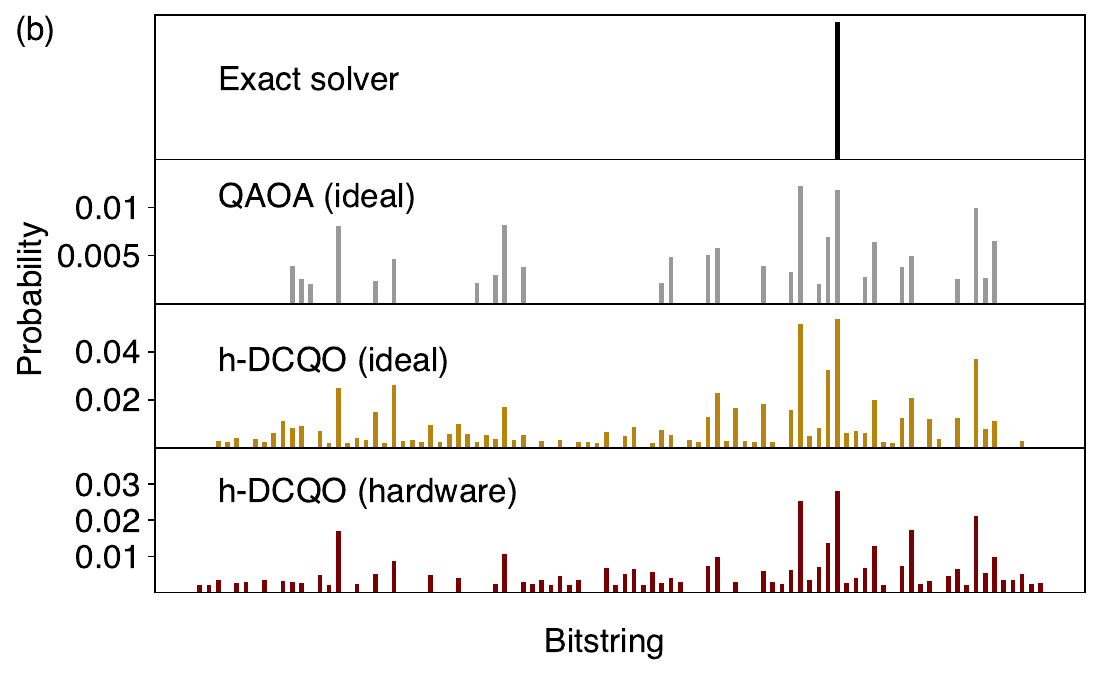}
    \caption{Comparing QAOA (ideal simulation) and h-DCQO (ideal simulation and IonQ implementation) for a 16-qubit subproblem of JSSP. Both QAOA and h-DCQO circuits have 240 MS gates. IonQ \texttt{Aria-1} is used for the demonstration and only the final circuit with optimized parameters is run on the hardware.
    (a) Energy distribution obtained from the final quantum circuit with optimized parameters. ARs are 50\%, 80\% and 63\% for QAOA (ideal), h-DCQO (ideal) and h-DCQO (hardware), respectively.
    (b) Probability distribution obtained from the final quantum circuit with optimized parameters.
    SPs are 1.2\%, 5.4\% and 2.8\% for QAOA (ideal), h-DCQO (ideal) and h-DCQO (hardware), respectively.
    }
    \label{fig:hdcqo_jssp}
\end{figure}

Now we zoom in on the performance comparison between h-DCQO and QAOA for the last instance in Fig.~\ref{fig:compare_jssp}(b). Here we report the best performance after iterating over 20 random initializations of QAOA circuit. 
In ideal simulation, h-DCQO is able to achieve $0.6\times$ (3.5$\times$) 
higher AR (SP) than QAOA, see Fig.~\ref{fig:hdcqo_jssp}. By constructing the variational ansatz using just the CD terms, we are able to obtain low-energy bitstrings with more probability; this leads to both higher AR and SP for h-DCQO algorithm.
To understand the impact of noise, we then run the final circuit, with optimal parameters, on IonQ's processor. Although the hardware noise, even after error mitigation, lowers the quality of the noiseless results, h-DCQO circuit still achieves superior performance than the ideal QAOA circuit. In particular, our demonstration yields a SP of 2.8\% and an AR of 63\%.
Moreover, we notice that the SP of DCQO is 3.4$\times$ more than that of QAOA, making DCQO superior. This is particularly interesting for NISQ applications because DCQO is a pure quantum algorithm and thus does not suffer from the challenges of variational quantum algorithms. 

\subsection{(h-)DCQO for TSP}
\label{subsec:tsp_results}
\begin{figure}
    \centering
    \includegraphics[width=\linewidth]{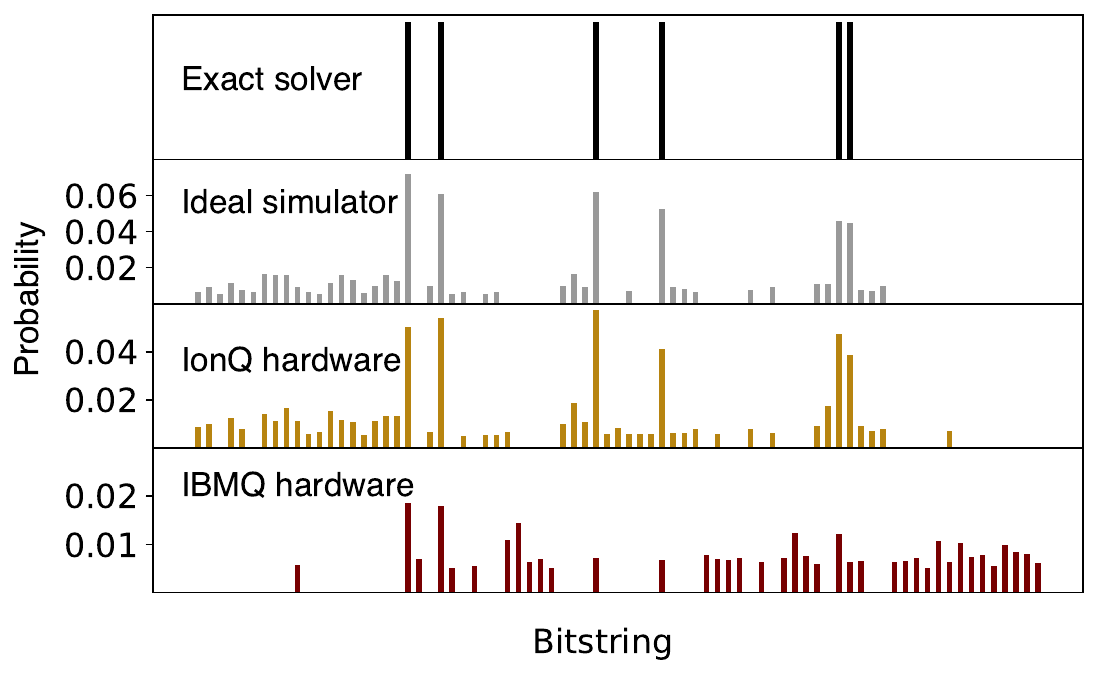}
    \caption{DCQO results for a 9-qubit TSP instance. For $T=0.2$, $N=2$ and a gate-cutoff threshold of 0.1~(sec. \ref{subsec:red}), the ideal DCQO circuit has 36 CX gates. Whereas the transpiled DCQO circuits run on IonQ~\texttt{Aria-2} and IBMQ~\texttt{ibmq\_guadalupe} have 36 MS gates and 66 CX gates, respectively.
    }
    \label{fig:dcqo_9tsp}
\end{figure}

\begin{figure}
    \centering
    \includegraphics[width=\linewidth]{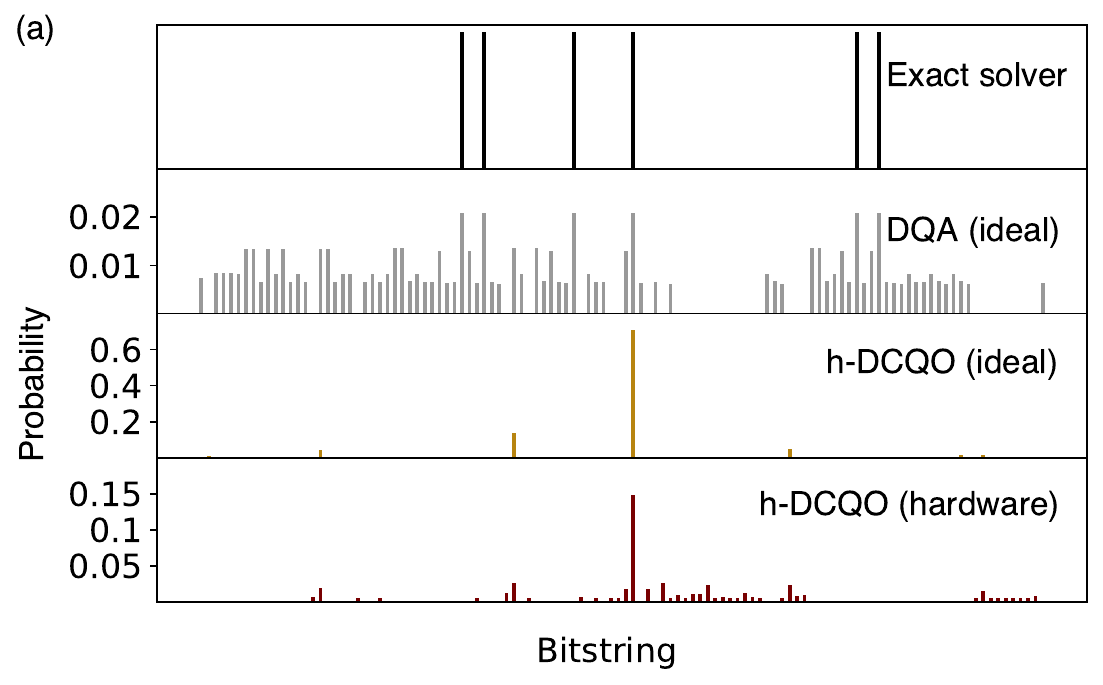}
    \includegraphics[width=\linewidth]{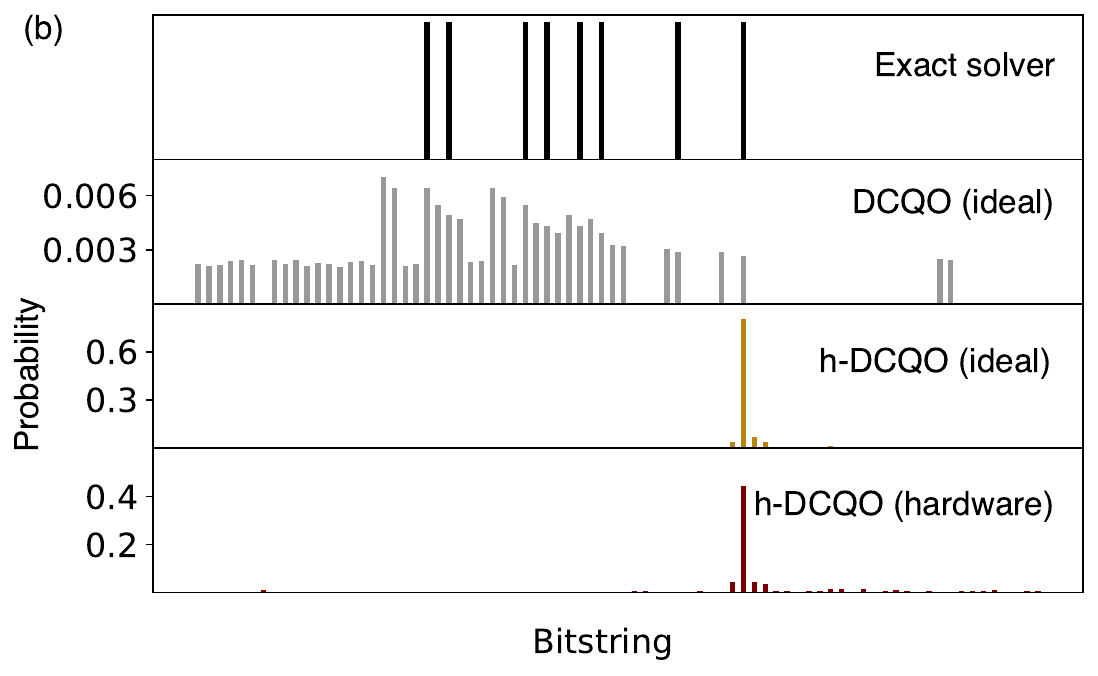}
    \caption{Hybrid DCQO vs DQA (DCQO) results for the 3-city (4-city) TSP instance. The h-DCQO ansatz has one layer consisting of $Y$ and $ZY$ terms.
    The hybrid optimization is performed on a noiseless simulation and only the final circuit with optimized parameters is run on the hardware.
    (a) 3-city TSP instance. 
    DQA simulation results in a very low contrast between the solution and infeasible bitstrings.
    IBMQ's \texttt{ibmq\_guadalupe} is used for the  demonstration and the transpiled h-DCQO circuit has 100 CX gates.
    (b) 4-city TSP instance. 
    DCQO simulation yields unsuccessful results.
    IonQ \texttt{Aria-1} is used for the  demonstration and the transpiled h-DCQO circuit has 48 MS gates. 
    }
    \label{fig:hdcqo_tsp}
\end{figure}

Having already established the superiority of counterdiabatic algorithms, i.e.\ DCQO and h-DCQO, over other NISQ algorithms, i.e.\ DQA and QAOA, for JSSP instances, we now present cloud-based demonstration results on using counterdiabatic algorithms for TSP instances. 
Although a three-city TSP instance is trivial as all possible paths are the shortest-distance paths, its mapping as a 9-qubit Ising Hamiltonian makes it moderately challenging to be solved on a NISQ hardware. 
Specifically, we study how DCQO fares against DQA in beating finite-time annealing and Trotter error in ideal simulation.
In Fig.~\ref{fig:dcqo_9tsp}, we observe that a noiseless simulation of DCQO is able to identify all six degenerate ground states, i.e.\ all possible solutions of the three-city TSP, with high probability and just two Trotter steps. 
On the other hand, DQA, even with $11\times$ more two-qubit gates, yields inadequate outcomes; see Fig.~\ref{fig:hdcqo_tsp}(a). Although the DQA results depend on the evolution time, a good quality solution would require a large number of Trotter steps, which makes this method infeasible for current quantum processors.
A successful DCQO result from \texttt{ibmq\_guadalupe} is possible for this TSP instance because of the low number of two-qubit gates in the transpiled circuit. 
Though the success probabilities and the quality of the solution are overall somewhat lower than the ones obtained from IonQ. 
For the latter demonstrations, the degeneracy is clearly restored~(Fig.~\ref{fig:dcqo_9tsp}).  Although preserving degeneracy is not a requirement for optimization problems, it can become useful when studying perturbations to an open system resulting in a breakdown of degeneracy. 

We adapt the existing h-DCQO ansatz~(\S\ref{subsec:hdcqo}) for solving the 3-city and 4-city TSP instances. In particular, we construct an ansatz by parameterizing all one-body terms independently and all two-body terms together with a single parameter; this leads to $N_\text{q}+1$ trainable parameters for a $N_\text{q}$-qubit circuit. Additionally, we omit all $YZ$ Pauli terms from the ansatz, which not only halves the required number of two-qubit (MS) gates but also results in a higher success probability.
For the 9-qubit TSP instance, we warm-start the parameters using a DCQO circuit. The variational training of the ten parameters took about 120 iterations for the simulated expectation value of the problem Hamiltonian Eq.~\eqref{eq:Hprob} to converge to the exact ground-state energy with high precision.
The final quantum circuit, with the optimal parameters, yields only one of the six degenerate ground states with a success probability of nearly 71\%, as seen in Fig.~\ref{fig:hdcqo_tsp}(a). This result is valid because all degenerate ground states in TSP encode the same travel path.
Upon running the final circuit on IBMQ's hardware, we are able to extract the desired solution with about 15\% probability. This is possible due to a low CX count of 100 and the high success probability in ideal simulation.

For the 4-city TSP, i.e.\ with 16 qubits, DCQO is unable to discern the eight degenerate solutions with significant probabilities as it could do for the 3-city TSP. On the other hand, h-DCQO successfully finds a solution with a very high probability; see Fig.~\ref{fig:hdcqo_tsp}(b). For this problem instance, the h-DCQO ansatz has 17 variational parameters, which we train by scanning over 20 different random initializations. Warm starting h-DCQO fails in this case because the chances of getting stuck in a local minima increase significantly with more parameters. 
From our hardware implementations, we observe that IonQ hardware, with only 5000 shots and error mitigation, yields the shortest-distance path with a high probability of 45\%. However, \texttt{ibm\_brisbane} fails for this instance because the error from the 180 two-qubit gates of the transpiled circuit results in a random output distribution.

\section{Conclusions and outlook}
\label{sec:conclusion}
We have used our DCQO algorithm and its hybrid version to solve 9- and 16-qubit Ising spin-glass problems on current quantum hardware. The problem instances are taken from physically motivated and industrially relevant combinatorial optimization problems, namely the traveling salesperson problem~(TSP) and a job-shop scheduling problem~(JSSP). Using ideal simulations, we show that DCQO consistently outperforms DQA and QAOA in terms of success probability and required number of two-qubit gates. 
The main challenge was to go beyond simple ideal simulation. Realizing a hardware implementation of an all-to-all connected problem, even for the problem sizes addressed here, quickly approaches the limits of currently available NISQ devices. Therefore, we implemented additional circuit modification strategies, including gate-cutoff and gate-reordering, to further reduce the depth of our compressed circuits. This is followed by an efficient transpilation for the given hardware, resulting in a successful implementation of our algorithms on NISQ devices.

We use both superconducting circuits and trapped ions to assess their feasibility in solving TSP and JSSP instances.
For the TSP, we successfully solve a 3-city problem instance on an IBM superconducting machine, as well as 3-city and 4-city instances on an IonQ trapped-ion machine. 
This is achieved by adapting our existing (h-)DCQO algorithm to create a low-depth circuit. These instances have degenerate solutions, which were successfully resolved by our DCQO algorithm both on simulation and hardware for the 3-city instance.
On the other hand, h-DCQC tends to amplify one of the degenerate solutions depending on its initial parameters.
We also investigated the simplest form of a non-trivial JSSP for a given laboratory application, which would require a QUBO representation with 360 variables. Since such a problem is still too large to be tackled with current quantum hardware, requiring access to a 360-qubit machine with high two-qubit gate fidelity and dense qubit connectivity, we solve only subproblems of size 16, requiring 16 qubits. We show that h-DCQO outperforms QAOA in terms of success probability and approximation ratio for all 16-qubit instances tested here. Due to resource limitations, we train the variational parameters classically using a noiseless simulator and run only the final circuit on the quantum hardware.


We have identified the algorithmic and the implementation challenges of our (hybrid) DCQO algorithm.
The success probability obtained from DCQO decreases exponentially with increasing number of qubits, making it currently infeasible for solving Ising problems beyond 20 qubits. On the other hand, h-DCQO can handle larger size problems due to the flexibility in the construction of the ansatz. In this algorithm, we combine the advantages of variational training of parameterized quantum circuit with counterdiabatic driving.
Although hybrid methods may be able to tackle larger problems and yield high approximation ratio, there is no guarantee of high success probability since the algorithm may suffer from the appearance of barren plateaus and local minima.

We can test the performance of other variants of counterdiabatic optimization algorithms for logistics scheduling use cases.
One such method is our recently  proposed bias-field DCQO, which demonstrated an empirical scaling advantage over DCQO and  successfully solved spin-glass problem instances with up to 100 qubits on NISQ hardware~\cite{GDS+24}. 
Additionally, it would be both interesting and challenging to investigate the performance of counterdiabatic optimization algorithms that include higher-order NC terms, as compared to DCQO, which has only first-order terms.
Yet another intriguing direction to explore is tensor-network-based circuit optimization of counterdiabatic Hamiltonians~\cite{ML23}.
A possible improvement of h-DCQO can be made using the meta-learning framework introduced in Ref.~\cite{chandarana2022meta}.
Alternative to the above digital algorithms, we can also make use of our digital-analog counterdiabatic optimization algorithms~\cite{KHG+24} to address industrially relevant use cases.

On the application side, possible next steps are as follows. First, one could apply our optimization algorithms to solve larger TSP instances by using an encoding scaling of~$n\log(n)$, i.e.\ using only $n\log(n)$ qubits instead of the currently used~$n^2$ qubits for a $n$-city TSP instance.
Second, one could explore our quantum optimization algorithms for a more-involved logistics use case, such as the vehicle routing problem. This has previously been tackled using quantum annealing and variational quantum algorithms~\cite{OVO22}. Using an improved bit-to-qubit encoding~\cite{LDT+23}, our DCQO algorithm has the potential to solve large instances without having to deal with the large measurement overhead of variational optimization methods. Third, we can use the large neighborhood search algorithm~\cite{PR19} in conjunction with DCQO to solve a simple, non-trivial JSSP instance. The performance of this hybrid digital-digital optimization algorithm can be compared to D-Wave's Hybrid Solver, i.e.\ a hybrid digital-analog algorithm, for the robot scheduling problem~\cite{LSJ+23}. 

\appendix
\section{Quantum optimization}~\label{appendix:qo}
In the following, we briefly outline quantum algorithms used to solving combinatorial optimization problems.
\subsection{Adiabatic quantum optimization}
Since adiabatic quantum computing (AQC) with non-stoquastic Hamiltonians is equivalent with the circuit model of quantum computing \cite{Aharonov2007}, any computational problem can in principle be treated with AQC. Therefore, AQC represents an innovative approach to tackle optimization problems using analog quantum processors~\cite{FGG+01}, an approach known as adiabatic quantum optimization (AQO). Central to AQC is the adiabatic theorem~\cite{FGGS00} which states that a physical system remains in an instantaneous eigenstate of its time-dependent Hamiltonian if a perturbation acts slow enough and the system exhibits a gap between the respective state and the rest of the Hamiltonian spectrum. In ideal AQC, a quantum system being in a ground state of a given initial Hamiltonian $H_\text{i}$, which is easy to prepare on a quantum device, is driven to the ground state of a final Hamiltonian $H_\text{p}$, which encodes the solution of the computational problem. The corresponding adiabatic dynamics is governed by a time-dependent adiabatic Hamiltonian defined as:
\begin{equation}\label{eq:Hanneal}
    H_\text{ad}(t) = \left(1 - \lambda(t)\right) H_\text{i} + \lambda(t) H_\text{p},
\end{equation}
where the scheduling function~$\lambda(t)\in [0,1]$ is continuously differentiable, and it guides the transition from the initial to the problem Hamiltonian. A common approach to $H_i$ in AQO is a transverse field
\begin{equation}
\label{eq:Hinit}
    H_\text{i} = -\sum_i \sigma_i^x,
\end{equation} 
where $\sigma^x$ is a Pauli matrix in Dirac notation, and whose ground state is a uniform superposition state in the computational basis. For the optimization problems treated in this work, we assume for the problem Hamiltonian $H_\text{p}$ an Ising spin-glass model according to Eq.~\eqref{eq:Hprob}. 
A transverse field appears to be a proper choice here because then $H_\text{i}$ and $H_\text{p}$ do not commute. 
The implementation challenge of AQO is to make the dynamics slow enough to avoid unwanted transitions to eigenstates different from the ground state while the adiabatic evolution is carried out in finite time. This leads to the concept of finite-time AQO which is typically referred to as quantum annealing. 

\subsection{Quantum annealing}
Quantum annealing is a meta-heuristic optimisation procedure that aims at solving a combinatorial optimisation problem using properties of quantum physics, and not requiring adiabatic evolution and universality~\cite{KN98}. This technique has been commercialized by D-Wave Systems, which owns the state-of-the-art quantum annealer with 5000+ qubits. The quantum annealing process is contingent on the preservation of quantum coherence. A prolonged annealing process, in comparison to the coherence time of the qubits, can diminish the quantum attributes of the system. Besides noisy qubits, one of the other inherent limitations of these annealers is the restricted connectivity between the qubits. This architectural constraint impedes the direct implementation of $H_\text{p}$, often compelling the use of embedding techniques, which comes at the cost of additional qubit requirement. Moreover, it is challenging to implement a non-stochastic problem Hamiltonian on these annealers. Thus, although D-Wave annealers are powerful for sampling near-optimal solutions for QUBO problems, it is not suitable for universal computation, i.e.\ not any computational problem is treatable on these devices. In contrast to that, digitized adiabatic quantum computing supports universality and error correction, and has been successfully implemented on a superconducting circuit with nine qubits~\cite{BSL+16}.

\subsection{Variational quantum algorithms}
Due to the above-mentioned challenges of analog quantum annealing and flexibility of gate-based quantum processors, quantum optimization algorithms are being developed for NISQ hardware. The two popular NISQ algorithms for combinatorial optimization are quantum approximate optimization algorithm~(QAOA) and variational quantum eigensolver~(VQE)~\cite{SGS+23}. Both are hybrid quantum-classical algorithms in a sense that a solution is reached variationally by making a classical and quantum computer work together in tandem. The two main components are parameterized quantum circuit, which acts typically on an equal superposition initial state to construct a parameterized quantum state, and classical optimizer. The expectation value of the problem Hamiltonian~\eqref{eq:Hprob} is estimated quantumly for the parameterized state, whose parameters are updated in an iterative manner using the classical optimizer such that the expectation value is minimized. 
The parameterized quantum circuit in QAOA has alternating layers of unitaries constructed from the problem Hamiltonian and a mixer Hamiltonian~\eqref{eq:Hinit}, respectively~\cite{FGG14}. A successful implementation of QAOA on digital hardware would require deep circuits, and hence infeasible on NISQ devices.
On the other hand, VQE for optimization employ single- and two-qubit gates available on a given hardware to construct the parameterized circuits~\cite{MBB+18}. This hardware-efficient VQE, although easily implementable on NISQ hardware, does not guarantee optimal solution.

\begin{widetext}
\section{Quantum hardware calibration data}~\label{appendix:cali}
\begin{table*}[h]
\label{tab:cali}
\begin{tabular}{ccc|cc|cccc}
\hline 
\hline
\multicolumn{3}{c|}{Hardware}  &
\multicolumn{2}{c|}{Fidelity} &
\multicolumn{4}{c}{Timing}\\
Name & Qubits & Connectivity  & 1Q gate & 2Q gate  & $T_1$ & $T_2$  & 1Q gate & 2Q gate  \\ \hline
\texttt{ibmq\_guadalupe} & 16  & Hexagonal & 0.9997 &0.9892  & 70 $\upmu$s& 88 $\upmu$s &27 ns & 394 ns\\ \hline
\texttt{Aria-1}  & 25  & All-to-all  & 0.9998 & 0.9831  & 100 s & 1 s& 135 $\upmu$s & 600 $\upmu$s \\ \hline
\texttt{Aria-2}   & 25  & All-to-all  & 0.9996 & 0.9775  & 10 s & 1.5 s & 135 $\upmu$s & 600 $\upmu$s  \\ \hline
\hline
\end{tabular}
\caption{Calibration data of the IBMQ and IonQ devices used in this work. We provide average values of all single-qubit (1Q) and two-qubit (2Q) gate fidelities during the time of our demonstration. Additionally, average of qubit coherence times, namely $T_1$ and $T_2$, and gate execution times are also reported. }
\end{table*}    
\end{widetext}

\bibliography{main.bib}

\end{document}